\documentclass[twocolumn,twocolappendix]{aastex702}

\usepackage{amsmath}
\usepackage{xcolor}

\newcommand{\kpolaris}{\textsc{KPolaris}}
\newcommand{\figorplaceholder}[2]{%
  \IfFileExists{#1}{\includegraphics[width=\linewidth]{#1}}{%
    \IfFileExists{figures/#1}{\includegraphics[width=\linewidth]{figures/#1}}{%
      \fbox{\parbox[c][43mm][c]{0.92\linewidth}{\centering
      \textcolor{red}{Missing figure: #2}}}}}}

\shorttitle{KPolaris: Polarized Radiative Transfer}
\shortauthors{Zhang et al.}

\begin{document}

\title{KPolaris: GPU-accelerated Polarized Radiative Transfer in General Relativity}

\author[orcid=0009-0008-9585-1807]{Zelin Zhang}
\affiliation{Institute of Fundamental Physics and Quantum Technology, \& School of Physical Science and Technology, Ningbo University, Ningbo, Zhejiang 315211, P. R. China}
\affiliation{Zhejiang Key Laboratory of Extreme Universe, \& BINGO Center, Ningbo University, Ningbo, Zhejiang 315211, China }
\email{zhangzelin1@nbu.edu.cn}

\author[orcid=0000-0003-4509-9705]{Bin Chen}
\email{chenbin1@nbu.edu.cn}
\affiliation{Institute of Fundamental Physics and Quantum Technology, \& School of Physical Science and Technology, Ningbo University, Ningbo, Zhejiang 315211, P. R. China}
\affiliation{Zhejiang Key Laboratory of Extreme Universe, \& BINGO Center, Ningbo University, Ningbo, Zhejiang 315211, China }
\affiliation{School of Physics, \& Center for High Energy Physics, Peking University, No.5 Yiheyuan Rd, Beijing 100871, P. R. China}
\correspondingauthor{Bin Chen}

\begin{abstract}
Polarized black-hole images carry information about both the magnetized plasma and the curved spacetime through which the radiation propagates. We present \kpolaris{}, an open-source GPU-accelerated general-relativistic polarized radiative-transfer code built on the Kokkos framework. It supports analytic plasma models and GRMHD simulation data, with both fast-light and slow-light radiative transfer. The code also provides diagnostics that connect image structure to the emitting plasma and to propagation effects along the rays. These include emissivity-weighted plasma maps, emission contributions from selected regions, selected-ray histories, and sensitivities of the Stokes parameters to plasma parameters. Numerical convergence tests and comparisons with an independent radiative-transfer code assess the accuracy of the calculations. GPU acceleration reduces the computational cost of image generation, facilitating multi-frequency and time-dependent studies of polarized black-hole images.
\end{abstract}

\keywords{black hole physics --- methods: numerical --- polarization ---
radiative transfer --- relativistic processes --- software: development}

\section{Introduction}\label{sec:introduction}

Horizon-scale observations provide a direct probe of magnetized plasma around black holes. The Event Horizon Telescope (EHT) has resolved the emission surrounding M87* and Sagittarius A* (Sgr A*), revealing bright rings with central brightness depressions consistent with black hole shadows \citep{EHTC2019,EHTC2022}. Polarimetric images show organized linear polarization in both sources \citep{EHTC2021VII,EHTC2024VII}, providing constraints on the magnetic field and physical conditions near the event horizon \citep{EHTC2021VIII,EHTC2024VIII}. Planned extensions of the EHT, in particular the next-generation EHT, aim to obtain images at multiple frequencies and to follow the evolution of the sources \citep{Johnson2023}. These prospects further increase the need for predictive models of polarized emission and variability.

Making these predictions requires connecting the dynamics of the accreting plasma to the polarized radiation received by a distant observer. General-relativistic magnetohydrodynamic (GRMHD) simulations provide the fluid and magnetic-field structure, while a prescription for the electron temperature and distribution function completes the plasma model needed to calculate the radiative coefficients \citep{Wong2022}.
For millimeter radiation from low-luminosity systems such as M87* and Sgr A*, these coefficients commonly describe synchrotron emission and absorption, together with Faraday rotation and conversion. These processes determine how polarized radiation is generated and modified along the ray, making the observed polarization sensitive to both the emitting regions and the intervening magnetized plasma \citep{Dexter2016,MoscibrodzkaGammie2018}.
General-relativistic radiative transfer (GRRT) follows this evolution along null geodesics, accounting for gravitational lensing, frequency shifts, and parallel transport to produce images, spectra, and light curves.

Numerical methods for polarized GRRT have been developed and tested through a range of independent implementations. These include \textsc{grtrans} \citep{Dexter2016}, \textsc{ipole} \citep{MoscibrodzkaGammie2018}, \textsc{Arcmancer} \citep{Pihajoki2018}, \textsc{BHOSS} \citep{Younsi2020}, \textsc{RAPTOR} \citep{Bronzwaer2020}, \textsc{Blacklight} \citep{White2022}, \textsc{GYOTO} 2.0 \citep{Aimar2024}, and \textsc{Coport} \citep{Huang2024}.
Although these codes solve the same physical transport problem, they use different polarization representations and integration methods. A common numerical challenge is the stiffness of the transfer equation at large optical or Faraday depth, which can be addressed using analytic updates or implicit integration.
Since these choices affect numerical accuracy, comparisons based on matched physical inputs and consistent polarization conventions are needed to benchmark GRRT calculations \citep{Gold2020,Prather2023} and to validate new implementations.

Applying these validated methods to systematic studies of black-hole images requires efficient calculations across many model realizations. Exploring a range of simulation times, viewing angles, observing frequencies, and electron prescriptions multiplies the number of transfer calculations \citep{EHTC2021VIII,Wong2022}.
For time-dependent studies, slow-light transfer further increases the computational demands by sampling the evolving plasma along each photon's trajectory to account for light-travel-time delays \citep{Bronzwaer2018}. Each image then requires access to multiple simulation snapshots, making data movement and memory management important alongside the integration itself.
The independence of pixel rays provides a natural route to accelerating these calculations through parallel execution on graphics processing units (GPUs), as demonstrated by \textsc{GRay} and \textsc{Odyssey} \citep{Chan2013,Pu2016} and by GPU-accelerated \textsc{ipole} for fully polarized transfer \citep{MoscibrodzkaYfantis2023}.
Efficient slow-light imaging therefore calls for combining parallel ray integration with the management of time-dependent simulation data.

Interpreting the images also requires understanding how radiation acquires its observed polarization. Emission reaching a single pixel can originate in spatially separated regions and be modified by different intervening absorption and Faraday effects. Studies of internal Faraday rotation and circular polarization illustrate the value of separating emission from propagation \citep{Ricarte2020,Ricarte2021}. Auxiliary plasma maps in \textsc{Blacklight} relate image structure to simulation variables \citep{White2022}, while differentiable calculations with \textsc{Jipole} quantify image responses to model parameters \citep{Motta2025,Motta2026}. These approaches motivate the development of diagnostics that can accompany efficient image generation.

We present \kpolaris{}\footnote{\url{https://github.com/zelinzh/KPolaris}}, an open-source code for polarized black-hole imaging and radiative diagnostics.
The code computes polarized radiative transfer through either analytic plasma models or GRMHD simulation data, with support for multiple observing frequencies and both fast-light and slow-light calculations. To carry out these calculations efficiently on CPUs and GPUs, we use Kokkos to maintain a common implementation across the two architectures \citep{Edwards2014,Trott2022}.
For slow-light calculations, this implementation also manages access to the sequence of simulation snapshots needed to sample the evolving plasma along each ray. Alongside the resulting images, the code provides plasma maps, spatial emission contributions, ray histories, and parameter sensitivities. These diagnostics allow the images to be examined in terms of where the radiation originates, how its polarization evolves during propagation, and how it responds to changes in the plasma model.

Section~\ref{sec:method} describes the numerical method, and Section~\ref{sec:models} introduces the plasma inputs. Sections~\ref{sec:verification} and \ref{sec:performance} present accuracy and performance tests. Section~\ref{sec:physical-diagnostics} demonstrates the selected-ray analysis, and Section~\ref{sec:conclusions} summarizes the results.

\section{Numerical Method}\label{sec:method}

\kpolaris{} traces null geodesics from a camera through an analytic plasma model or GRMHD data and integrates polarized transfer toward the observer. We use a covariant formulation of polarized radiative transfer
in a parallel-transported polarization basis \citep{Pihajoki2018}, metric signature $(-,+,+,+)$, and geometrized units $G=c=1$ for the ray geometry. Radiation coefficients are evaluated in physical units using the models in Section~\ref{sec:models}.

\subsection{Spacetime and observer}
\label{sec:geometry-observer}

Kerr spacetime is implemented in Boyer--Lindquist, Cartesian and spherical Kerr--Schild, modified Kerr--Schild (MKS; \citealt{Gammie2003}), and funky modified Kerr--Schild (FMKS; \citealt{Wong2022}) coordinates. Supported simulation inputs are traced in their native coordinates. Analytic metric derivatives or specialized connection contractions evolve the photon momentum and polarization basis without storing the full connection array per GPU thread.

The camera is specified by a position, four-velocity, and orthonormal tetrad. Pinhole rays launch from a common event with different directions \citep{Gold2020,MoscibrodzkaGammie2018}; plane-parallel rays start on an image plane with a common direction \citep{Pihajoki2018}. Two screen vectors $e_A^\mu$, $A=1,2$, are orthogonal to the photon direction and observer velocity. The vertical axis follows the projected black hole spin axis, and the screen is right-handed with respect to physical photon propagation. Stokes parameters are reported in this camera basis, with $\mathrm{EVPA}=\tfrac12\operatorname{atan2}(U,Q)$ and the sign of $V$ inherited from the synchrotron coefficient conventions \citep{Pandya2016,Marszewski2021}. 
% Transformations to sky conventions are applied explicitly in post-processing \citep{Prather2023}.

\subsection{Ray tracing and parallel transport}
\label{sec:doublepass}

Let $\lambda$ be an affine parameter and $k^\mu=dx^\mu/d\lambda$ the photon wave vector. The ray and the parallel-transported screen basis vectors satisfy
\begin{align}
 \frac{d x^\mu}{d\lambda} &= k^\mu, \label{eq:ray-position}\\
 \frac{d k^\mu}{d\lambda} &=-\Gamma^\mu_{\alpha\beta}k^\alpha k^\beta,
 \label{eq:ray-momentum}\\
 \frac{d e_A^\mu}{d\lambda} &=-\Gamma^\mu_{\alpha\beta}k^\alpha e_A^\beta.
 \label{eq:screen-transport}
\end{align}
The geometrical equations use fourth-order Runge--Kutta integration with step doubling for local error control. Parallel transport carries the screen basis, leaving the local plasma terms to govern polarized transfer in that basis.

Independent pixel rays are parallelized with Kokkos \citep{Edwards2014,Trott2022}. Following the bidirectional strategy of GPU-accelerated ipole \citep{MoscibrodzkaYfantis2023}, Pass A traces from the camera to the inner boundary or domain exit with negative affine-parameter increments. Pass B retains the terminal wave vector and screen basis and uses positive increments to return toward the camera while solving the transfer equation. The wave vector is future directed in both passes. Recomputing the geometry avoids storing each sampled trajectory, giving persistent storage proportional to the number of pixels for fixed frequencies and outputs.

At the camera, the returned screen basis is matched to the original camera axes through their overlap. Closure, null, and screen-basis constraints are recorded as diagnostics. Full trajectories are retained only for selected rays. Integration controls are listed in Appendix~\ref{app:controls}.

\subsection{Invariant polarized transfer}
\label{sec:polarized-transfer}

The invariant Stokes vector is
\begin{equation}
 \boldsymbol{\mathcal S}
 =\frac{1}{\nu^3}(I_\nu,Q_\nu,U_\nu,V_\nu)^{\mathsf T},
\end{equation}
where $\nu$ is the photon frequency measured in the local frame. With the metric signature $(-,+,+,+)$, the photon frequency measured by an observer with four-velocity $u^\mu$ is proportional to $-u_\mu k^\mu$. The frequency in the fluid frame is therefore
\begin{equation}\label{eq:local-frequency}
 \nu=\nu_{\rm obs}
 \frac{-u_\mu k^\mu}{-u_{{\rm obs}\,\mu}k^\mu_{\rm obs}},
\end{equation}
where $u^\mu$ and $u_{\rm obs}^\mu$ are the fluid and camera four-velocities.

In the transported basis, the transfer equation is
\begin{equation}\label{eq:transfer}
 \frac{d\boldsymbol{\mathcal S}}{d\lambda}
 =\boldsymbol{\mathcal J}-\mathbf K\boldsymbol{\mathcal S},
\end{equation}
with propagation matrix
\begin{equation}\label{eq:propagation-matrix}
 \mathbf K=
 \begin{pmatrix}
 \mathcal A_I & \mathcal A_Q & \mathcal A_U & \mathcal A_V\\
 \mathcal A_Q & \mathcal A_I & \mathcal R_V & -\mathcal R_U\\
 \mathcal A_U & -\mathcal R_V & \mathcal A_I & \mathcal R_Q\\
 \mathcal A_V & \mathcal R_U & -\mathcal R_Q & \mathcal A_I
 \end{pmatrix}.
\end{equation}
Here $\boldsymbol{\mathcal J}$ describes emission, $\mathcal A_a$ absorption, and $\mathcal R_V$ and $\mathcal R_{Q,U}$ Faraday rotation and conversion. The coefficients per unit affine parameter are related to their
fluid-frame values by
\begin{equation}\label{eq:affine-coefficients}
 \boldsymbol{\mathcal J}=\frac{\ell_\lambda}{\nu^3}\boldsymbol j_\nu,
 \qquad
 \mathcal A_a=\ell_\lambda\alpha_{\nu,a},
 \qquad
 \mathcal R_a=\ell_\lambda\rho_{\nu,a},
\end{equation}
where $\ell_\lambda=d\ell/d\lambda$ is the fluid-frame path length per unit affine parameter, including the conversion to the length units used by the physical coefficients. In consistent geometrized units, $\ell_\lambda=-u_\mu k^\mu$. The transported screen is projected into the fluid rest frame, and the emissivity vector and propagation matrix are rotated from the local magnetic-field basis into that screen. We use the standard basis transformations described by \citet{Pihajoki2018}. The radiation coefficient prescriptions are specified in Section~\ref{sec:models}.

The coefficients are held constant within each radiation substep. We split the propagation matrix into absorption and Faraday parts, $\mathbf K=\mathbf K_{\rm A}+\mathbf K_{\rm F}$. The Stokes vector is advanced through a Faraday half-step, an emission--absorption full-step, and a second Faraday half-step
\begin{equation}\label{eq:strang-transfer}
 \boldsymbol{\mathcal S}_{n+1}
 =\mathsf F_{\Delta\lambda/2}
 \left[
  \mathsf A_{\Delta\lambda}
  \left[
   \mathsf F_{\Delta\lambda/2}
   \left(\boldsymbol{\mathcal S}_n\right)
  \right]
 \right].
\end{equation}
Here $\mathsf F_h$ and $\mathsf A_h$ denote the analytic constant-coefficient updates over an affine interval $h$ for Faraday rotation and conversion, and for emission and absorption, respectively. This symmetric Strang splitting is second-order accurate for constant coefficients \citep{Strang1968}.

Analytic substeps handle stiff absorption and Faraday terms, while step limits control splitting errors and coefficient variation along the ray. Geometrical tolerance, maximum geometrical and radiation steps, and
absorption and Faraday depth limits constrain the integration separately. The calculations assume zero incident radiation at the far endpoint.

\subsection{Multi-frequency and slow-light transfer}
\label{sec:frequency-time}

In the null-geodesic approximation, multiple frequencies share a trajectory but retain separate Stokes vectors, coefficients, and radiation-step controls. Equation~\eqref{eq:local-frequency} gives the fluid-frame frequency at each event.

Fast-light calculations hold the plasma fixed in one snapshot. Slow-light calculations instead sample its evolution at the coordinate times encountered along the geodesic. At each event, \kpolaris{} samples the two bracketing snapshots and interpolates the fluid quantities linearly in time before evaluating the transfer coefficients. The snapshot sequence must therefore cover the full range of times sampled along the rays.

To process sequences larger than GPU memory, snapshots are streamed through a bounded resident window, with optional host-side prefetching. \kpolaris{} also supports the dense-dump-codec (DDC)\footnote{\url{https://github.com/zelinzh/dense-dump-codec}} compressed format \citep{zhang_2026_22845273}, which reduces storage requirements. After decoding, the data are processed using the same plasma and radiative-transfer calculations as native inputs.

\subsection{Radiative diagnostics and parameter sensitivities}
\label{sec:analysis-outputs}
\label{sec:response-definitions}

Optional outputs include emissivity-weighted plasma maps, optical and Faraday depths, and selected-ray histories. Spatial emission contributions are obtained by restricting the source while retaining absorption and Faraday effects along the full ray. Contributions from an exhaustive partition sum to the complete Stokes image for zero incident radiation. Definitions and checks are given in Appendix~\ref{app:diagnostic-definitions}.

Parameter sensitivities describe small radiative perturbations of a given flow. For $q=\ln f$, density normalization scales $n_e\mapsto fn_e$ and $B\mapsto f^{1/2}B$ at fixed $\Theta_e$, temperature scales $\Theta_e\mapsto f\Theta_e$ at fixed density and field, and magnetic strength scales $B\mapsto fB$ at fixed density and temperature. Geometry, velocity, field direction, and the baseline emitting mask remain fixed. The temperature perturbation acts on the final $\Theta_e$, not on $R_{\rm high}$, and these responses do not include dynamical readjustment of the flow.

Differentiating Equation~\eqref{eq:transfer} gives
\begin{equation}\label{eq:tangent}
 \frac{d\boldsymbol{\mathcal S}_q}{d\lambda}
 =-\mathbf K\boldsymbol{\mathcal S}_q
  +\boldsymbol{\mathcal J}_q-\mathbf K_q\boldsymbol{\mathcal S},
 \qquad
 \boldsymbol{\mathcal S}_q=\frac{\partial\boldsymbol{\mathcal S}}{\partial q}.
\end{equation}
The derivative is evaluated at $f=1$ with zero incident sensitivity. At each radiation substep, we evaluate the local semi-analytic transfer update at $f=e^{\pm h}$ using the same incident Stokes vector. Their centered difference gives the local response, which is added to the incoming sensitivity propagated by the unperturbed transfer operator. Responses associated with emission, absorption, rotation, and conversion sum to the total derivative. Appendix~\ref{app:response-checks} compares these responses with finite differences of complete transfer solutions.

\section{Models and Inputs}\label{sec:models}

\kpolaris{} supports both analytic plasma models and GRMHD simulation data. The implemented analytic models include a radiatively inefficient accretion flow (RIAF) and an equilibrium magnetized torus. The RIAF model combines parameterized density and electron-temperature profiles with a magnetic-field prescription and a velocity field interpolating between Keplerian rotation and free fall \citep{Broderick2016,PuBroderick2018}. The magnetized torus model describes a stationary, axisymmetric torus with a toroidal magnetic field \citep{Komissarov2006}. GRMHD simulation data can be read from the iharm3D HDF5 format \citep{Prather2021} and the native formats of BHAC \citep{Porth2017}, KHARMA \citep{Prather2025}, and AthenaK \citep{Stone2026}. KHARMA inputs retain their native modified Kerr--Schild representation, while AthenaK inputs retain the mesh-refinement structure in Cartesian Kerr--Schild coordinates. 

GRMHD density and internal energy are converted to physical units using the black hole mass and density normalization factor. For rest-mass density $\rho$ and internal energy density $u$, we define $p_{\rm gas}=(\Gamma-1)u$, $p_{\rm mag}=b^2/2$, $\beta=p_{\rm gas}/p_{\rm mag}$, and $\sigma=b^2/\rho$, where $b^2=b^\mu b_\mu$ and $\Gamma$ is the simulation adiabatic index.

The proton-to-electron temperature ratio, $R=T_p/T_e$, is prescribed as a function of plasma beta following \citet{Moscibrodzka2016},
\begin{equation}\label{eq:electron-ratio}
 R(\beta)=
 \frac{R_{\rm low}+R_{\rm high}(\beta/\beta_{\rm crit})^2}
 {1+(\beta/\beta_{\rm crit})^2}.
\end{equation}
The parameters $R_{\rm low}$ and $R_{\rm high}$ set the limiting temperature ratios at low and high plasma beta, while $\beta_{\rm crit}$ sets the transition between them. For an electron--proton plasma, partitioning the internal energy between the two species gives the dimensionless electron temperature $\Theta_e=k_{\rm B}T_e/(m_ec^2)$ \citep{Wong2022},
\begin{equation}\label{eq:electron-temperature}
 \Theta_e=
 \frac{m_p}{m_e}
 \frac{(\gamma_e-1)(\gamma_p-1)}
 {(\gamma_p-1)+(\gamma_e-1)R(\beta)}
 \frac{u}{\rho},
\end{equation}
where $u/\rho$ is expressed in code units. We adopt $\gamma_e=4/3$ and $\gamma_p=5/3$ for the electron and proton energy partition. These indices specify the radiative post-processing prescription, whereas $\Gamma$ is the adiabatic index used in the GRMHD simulation. 

The radiation module supports thermal, power-law, and $\kappa$ electron distributions. For thermal electrons, the synchrotron emissivities are evaluated using the fits of \citet{Pandya2016} or \citet{Dexter2016}, and the absorptivities follow from Kirchhoff's law. The thermal Faraday coefficients follow \citet{Shcherbakov2008}, with the conversion-coefficient correction given by \citet{Dexter2016}. For power-law and $\kappa$ distributions, polarized emissivity and absorptivity fits are provided by \citet{Pandya2016,Marszewski2021}. The $\kappa$-distribution Faraday coefficients use the fits of \citet{Marszewski2021}. The fits are used within their published domains of validity.

\section{Numerical Accuracy}\label{sec:verification}

We test integration convergence, compare images with ipole, and assess the accuracy of the settings used in the performance measurements.

\subsection{Integration convergence and image metrics}
\label{sec:convergence-metrics}

For constant coefficients, the separate emission, absorption, and pure Faraday-rotation tests reproduce their analytic solutions to floating-point precision. When absorption and Faraday terms act together, step refinement gives second-order convergence, consistent with the symmetric composition in Section~\ref{sec:polarized-transfer}. Tests with spatially varying coefficients also exhibit second-order convergence. The geodesics and parallel-transported screen vectors converge at fourth order under fixed-step refinement, as expected for the geometrical integrator. Checks of the selected-ray transfer reconstruction and parameter sensitivities are given in Appendices~\ref{app:cp-accounting} and \ref{app:response-checks}.

For complete images, we quantify differences using the normalized mean squared error (NMSE), following \citet[][Equation~19]{Prather2023}. For each Stokes component $S\in\{I,Q,U,V\}$,
\begin{equation}\label{eq:image-nmse}
 \mathrm{NMSE}(S)=
 \frac{\sum_p(S_p-S_{{\rm ref},p})^2}
      {\sum_p S_{{\rm ref},p}^2},
\end{equation}
where the sum extends over all image pixels. Each component is normalized by its own reference image, and $\sqrt{\mathrm{NMSE}}$ is the relative $L_2$ difference. We also compare total flux and the net linear polarization fraction, $m_{L,\rm net}=\sqrt{(\sum_p Q_p)^2+(\sum_p U_p)^2}/\sum_p I_p$.

\subsection{Independent-code image comparisons}
\label{sec:ipole-comparison}

We compare \kpolaris{} with ipole using the same plasma prescription, observing frequency, camera, pixel grid, and radiation-coefficient prescriptions. The output Stokes vectors are expressed in a common screen convention. Comparisons use the original pixel values, without image registration, fitted flux normalization, or smoothing. All images have $256^2$ pixels and are calculated at 230 and 345 GHz; model parameters and integration controls are listed in Appendices~\ref{app:samples} and \ref{app:controls}.

The fast-light tests cover an analytic RIAF, a public KHARMA MAD snapshot with $a=0.9375$ \citep{Dhruv2025}, and a native AthenaK MAD snapshot. This selection tests the image calculation for a continuous analytic plasma and for simulation data in both modified spherical and Cartesian Kerr--Schild coordinates. Table~\ref{tab:ipole-nmse} reports all four Stokes components. The fast-light intensity NMSEs are below $2\times10^{-5}$, and the largest component NMSE is $1.10\times10^{-4}$.

\begin{figure*}[!t]
 \centering
 \figorplaceholder{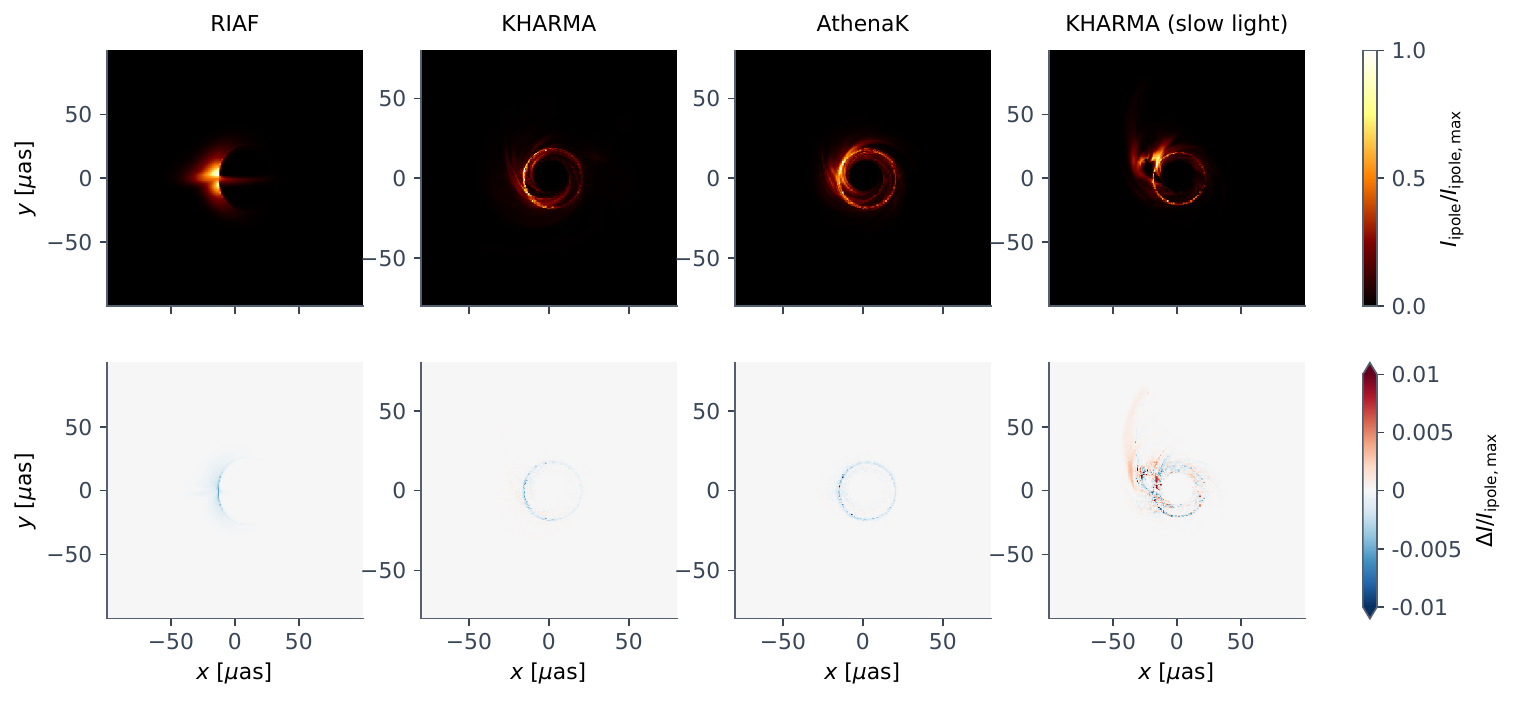}{Intensity images and residuals relative to ipole.}
 \caption{Intensity-image comparison with ipole at $256^2$ and 230 GHz. Columns show the fast-light RIAF, KHARMA, and AthenaK tests, followed by the slow-light KHARMA MAD test with $a=0.5$. Top: ipole intensity divided by its peak value in each image. Bottom: signed residuals, $(I_{\rm KPolaris}-I_{\rm ipole})/I_{\rm ipole,max}$. Both rows use linear color scales. The comparison uses no image registration, fitted normalization, or smoothing. Component-wise NMSEs at both observing frequencies are given in Table~\ref{tab:ipole-nmse}.}
 \label{fig:comparison}
\end{figure*}

The slow-light test uses the KHARMA MAD sequence with $a=0.5$ and cadence of approximately $0.1M$, with fluid interpolation as described in Section~\ref{sec:frequency-time}. Its largest component NMSE is $7.60\times10^{-4}$. Total-intensity flux differences relative to ipole are $0.113\%$ and $0.126\%$ at 230 and 345 GHz, respectively. The largest net linear polarization difference is $0.0787$ percentage points.  Figure~\ref{fig:comparison} shows the 230 GHz intensity residuals.

\begin{deluxetable*}{lrrrr}[!t]
\tablecaption{Stokes-image NMSE relative to ipole at $256^2$}\label{tab:ipole-nmse}
\tablewidth{0pt}
\tabletypesize{\footnotesize}
\tablehead{\colhead{Model and frequency} &
\colhead{NMSE(I)} & \colhead{NMSE(Q)} & \colhead{NMSE(U)} & \colhead{NMSE(V)}}
\startdata
RIAF, 230 GHz & $9.27\times10^{-6}$ & $6.12\times10^{-5}$ & $6.50\times10^{-5}$ & $3.72\times10^{-6}$ \\
RIAF, 345 GHz & $1.12\times10^{-5}$ & $8.03\times10^{-5}$ & $5.36\times10^{-5}$ & $3.32\times10^{-6}$ \\
KHARMA, 230 GHz & $1.56\times10^{-5}$ & $1.03\times10^{-4}$ & $6.18\times10^{-5}$ & $2.45\times10^{-5}$ \\
KHARMA, 345 GHz & $1.85\times10^{-5}$ & $1.10\times10^{-4}$ & $6.68\times10^{-5}$ & $2.46\times10^{-5}$ \\
AthenaK, 230 GHz & $1.39\times10^{-5}$ & $7.12\times10^{-5}$ & $6.70\times10^{-5}$ & $1.53\times10^{-5}$ \\
AthenaK, 345 GHz & $1.81\times10^{-5}$ & $6.97\times10^{-5}$ & $7.61\times10^{-5}$ & $2.51\times10^{-5}$ \\
KHARMA (slow light), 230 GHz & $1.27\times10^{-4}$ & $2.64\times10^{-4}$ & $5.35\times10^{-4}$ & $7.60\times10^{-4}$ \\
KHARMA (slow light), 345 GHz & $1.64\times10^{-4}$ & $2.28\times10^{-4}$ & $4.52\times10^{-4}$ & $5.54\times10^{-4}$ \\
\enddata
\tablecomments{The first six rows are fast-light comparisons, and the last two use the time-dependent KHARMA MAD sequence. Each Stokes component is normalized by its own ipole reference image, as in Equation~\eqref{eq:image-nmse}. The ipole step-size parameter is $\epsilon=3\times10^{-4}$ for fast light and $5\times10^{-4}$ for slow light. The corresponding \kpolaris{} controls are listed in Table~\ref{tab:integration-controls}, and model and camera parameters in Table~\ref{tab:comparison-inputs}.}
\end{deluxetable*}

\subsection{Accuracy at the performance settings}
\label{sec:accuracy-cost}

\begin{figure*}[!t]
 \centering
 \figorplaceholder{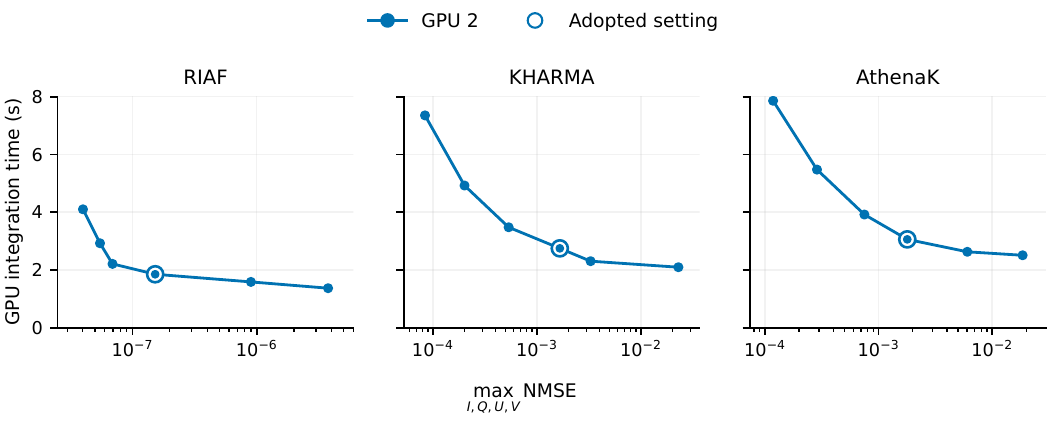}{Image error versus GPU integration time.}
 \caption{Fast-light integration time versus image error on an Nvidia RTX 5090 at $256^2$ pixels and 230 GHz. The geometrical tolerance varies from $10^{-13}$ to $10^{-8}$, with all other controls fixed as in Table~\ref{tab:daily-performance}. The open ring marks the adopted tolerance of $10^{-10}$. Times include geodesic integration and polarized transfer. Points show the median execution times from three runs. The error is the largest component NMSE relative to a more finely integrated \kpolaris{} image of the same model.}
 \label{fig:accuracy-cost}
\end{figure*}

To assess the performance settings separately from the refined ipole comparisons, Figure~\ref{fig:accuracy-cost} plots integration time against image error on an Nvidia RTX 5090. We vary the geometrical tolerance while holding the other controls fixed. The reference is a more finely integrated \kpolaris{} image of the same model, and the error is the largest NMSE among the four Stokes components.

At the adopted tolerance of $10^{-10}$, the maximum component NMSE is $1.53\times10^{-7}$ for the RIAF and below $1.8\times10^{-3}$ for both GRMHD inputs (Table~\ref{tab:daily-performance}). These comparisons measure integration accuracy for fixed plasma inputs, not convergence with GRMHD grid resolution. 
% The separate slow-light throughput test has its own first-frame accuracy reference in Table~\ref{tab:slowlight-throughput}.

\section{Performance}\label{sec:performance}

We compare Serial, OpenMP, and CUDA execution of the same double-precision implementation. CPU $n$ and GPU $n$ denote the laptop, workstation, and server processor groups in Table~\ref{tab:hardware}. Reported image-generation times include input preparation, data transfer, integration, and output.

\subsection{Fast-light image generation}
\label{sec:fastlight-performance}

Table~\ref{tab:platform-times} compares complete image-generation times at $256^2$ pixels and 230 GHz for an analytic RIAF and two GRMHD inputs. The timings include initialization, local input preparation, transfer to the GPU when applicable, ray integration, polarized transfer, and HDF5 image output. All processor groups use identical inputs and numerical controls. The accuracy associated with the adopted settings is characterized in Section~\ref{sec:accuracy-cost}, with the settings and representative errors given in Table~\ref{tab:daily-performance}.

\begin{deluxetable*}{llrrr}[!t]
\tablecaption{Complete fast-light image-generation times at $256^2$ and 230 GHz}
\label{tab:platform-times}
\tablewidth{0pt}
\tabletypesize{\footnotesize}
\tablehead{\colhead{Processor group} & \colhead{Model} &
\colhead{CPU, 1 thread (s)} & \colhead{CPU, all threads (s)} &
\colhead{GPU (s)}}
\startdata
1: Laptop & RIAF & 155.09 & 16.86 & 12.91 \\
 & KHARMA & 242.32 & 26.40 & 18.49 \\
 & AthenaK & 226.23 & 29.10 & 23.49 \\
2: Workstation & RIAF & 100.66 & 5.95 & 2.13 \\
 & KHARMA & 160.90 & 8.49 & 3.19 \\
 & AthenaK & 130.30 & 8.22 & 5.09 \\
3: Server & RIAF & 206.40 & 1.39 & 1.14 \\
 & KHARMA & 326.04 & 2.36 & 1.79 \\
 & AthenaK & 269.73 & 3.64 & 4.42 \\
\enddata
\tablecomments{Medians of three runs using the controls in
Table~\ref{tab:daily-performance}. The multithreaded CPU calculations use
all available hardware threads. Processor specifications are given in
Table~\ref{tab:hardware}. All times cover complete image generation.}
\end{deluxetable*}

GPU execution reduces the time relative to the corresponding single-threaded CPU calculation by factors of approximately 10--182 across these tests. Complete images take 2.13--5.09 s on the workstation GPU and 1.14--4.42 s on the server GPU. The comparison with fully threaded CPUs shows smaller, workload-dependent differences. GPU execution is faster in eight of the nine cases, while the server CPU completes the AthenaK calculation faster than its paired GPU. 

\begin{figure*}[!t]
 \centering
 \figorplaceholder{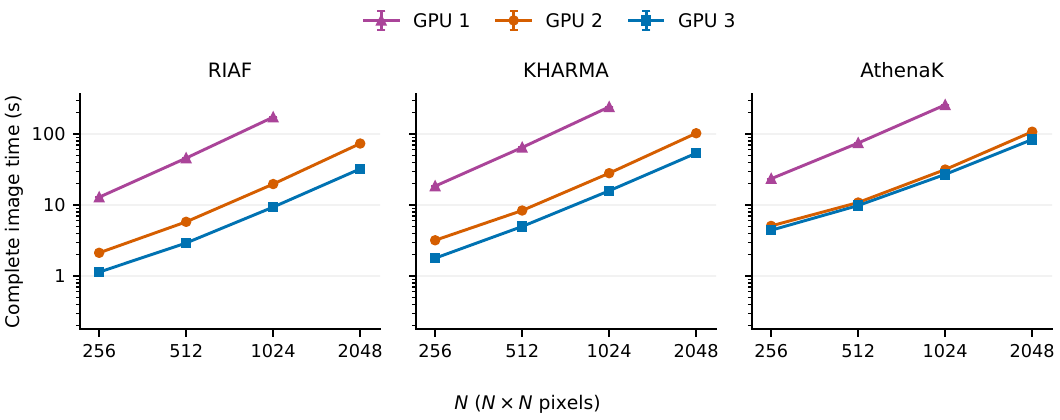}{Complete image time versus resolution.}
 \caption{Complete fast-light image-generation time versus resolution at
 230 GHz. All GPUs use the same source code, model inputs, and integration
 controls from Table~\ref{tab:daily-performance}. Points give medians of
 three runs, and bars span their measured times. GPU identifiers refer to
 Table~\ref{tab:hardware}.}
 \label{fig:performance}
\end{figure*}

Figure~\ref{fig:performance} shows the resolution dependence at fixed
integration settings. The time per pixel decreases for larger images
as fixed costs are shared among more rays.

\subsection{GPU slow-light throughput}
\label{sec:slowlight-performance}

We next measure the cost of producing slow-light polarized image sequences on GPU 2 using the KHARMA MAD simulation with $a=+0.5$. Each image has $256^2$ pixels at 230 GHz, with a nominal snapshot spacing of $0.5M$ and an image spacing of $0.5M$. The batch reuses the Pass A endpoints and transported screen bases across frames. Pass B is integrated separately for each frame, sampling the evolving fluid at the corresponding times. Inputs are stored locally with HDF5 compression removed losslessly before timing, and input file caches are cleared before each run. The elapsed times include process startup, input reading, GPU upload, integration, and image output. Table~\ref{tab:slowlight-throughput} gives medians of three independent runs for the same 16-image sequence under two GPU memory configurations. Numerical controls are recorded in Appendix~\ref{app:performance-setup}.

\begin{deluxetable*}{lrrr}[!t]
\tablecaption{Slow-light sequence throughput on an RTX 5090}
\label{tab:slowlight-throughput}
\tablewidth{0pt}
\tabletypesize{\footnotesize}
\tablehead{\colhead{Configuration} &
\colhead{Peak GPU memory (GiB)} & \colhead{Batch time (s)} &
\colhead{Time per image (s)}}
\startdata
Minimum residency & 5.4 & 54.00 & 3.38 \\
Intermediate residency & 15.1 & 39.45 & 2.47 \\
\enddata
\tablecomments{The two configurations process the same 16-image sequence with
identical input data, physical model, and integration settings,
but different snapshot-residency settings. The final column is the total
elapsed time divided by 16. Corresponding Stokes arrays are bitwise identical between configurations. For the first frame, comparison with a reference computed using stricter integration settings gives $\mathrm{NMSE}(I)=6.90\times10^{-4}$.}
\end{deluxetable*}

The two configurations produce images at 3.38 and 2.47 s per frame with peak GPU memory use of 5.4 and 15.1 GiB, respectively. Using nearly the full 32 GiB device did not improve the 16-image batch time. At intermediate residency, doubling the batch from 8 to 16 images leaves the total time near 39.5 s, demonstrating the benefit of batching. These amortized times are comparable in order of magnitude to the fast-light times, although the latter describe individual images with warm input caches.

\subsection{Cost of radiative diagnostics}
\label{sec:diagnostic-performance}

For the KHARMA test at $400^2$ on GPU 2, radial emission contributions in 16 bins and selected-region contributions increase the elapsed time by 12\% and 18\%, respectively. The electron-temperature response increases it by 78\%. Each value is based on two paired repetitions with otherwise identical integration settings. 

\section{Selected-Ray Analysis}\label{sec:physical-diagnostics}

\begin{figure*}[!t]
 \centering
 \begin{minipage}{0.80\linewidth}
  \figorplaceholder{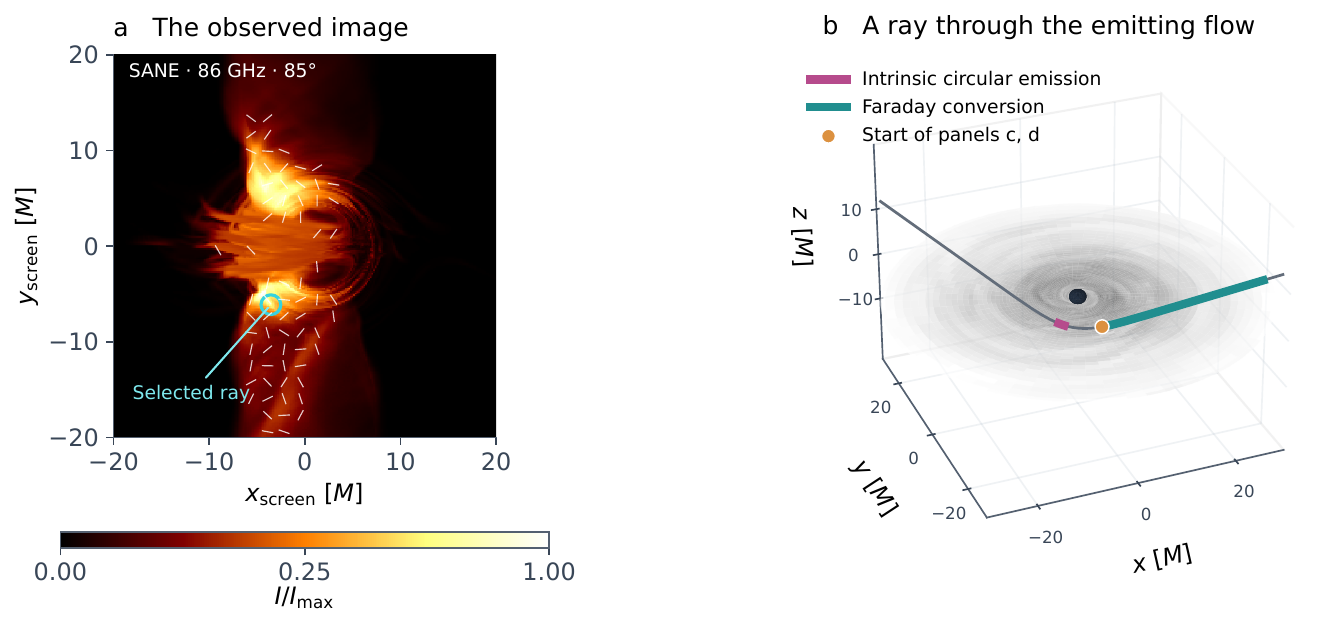}{Image and selected geodesic.}
  \figorplaceholder{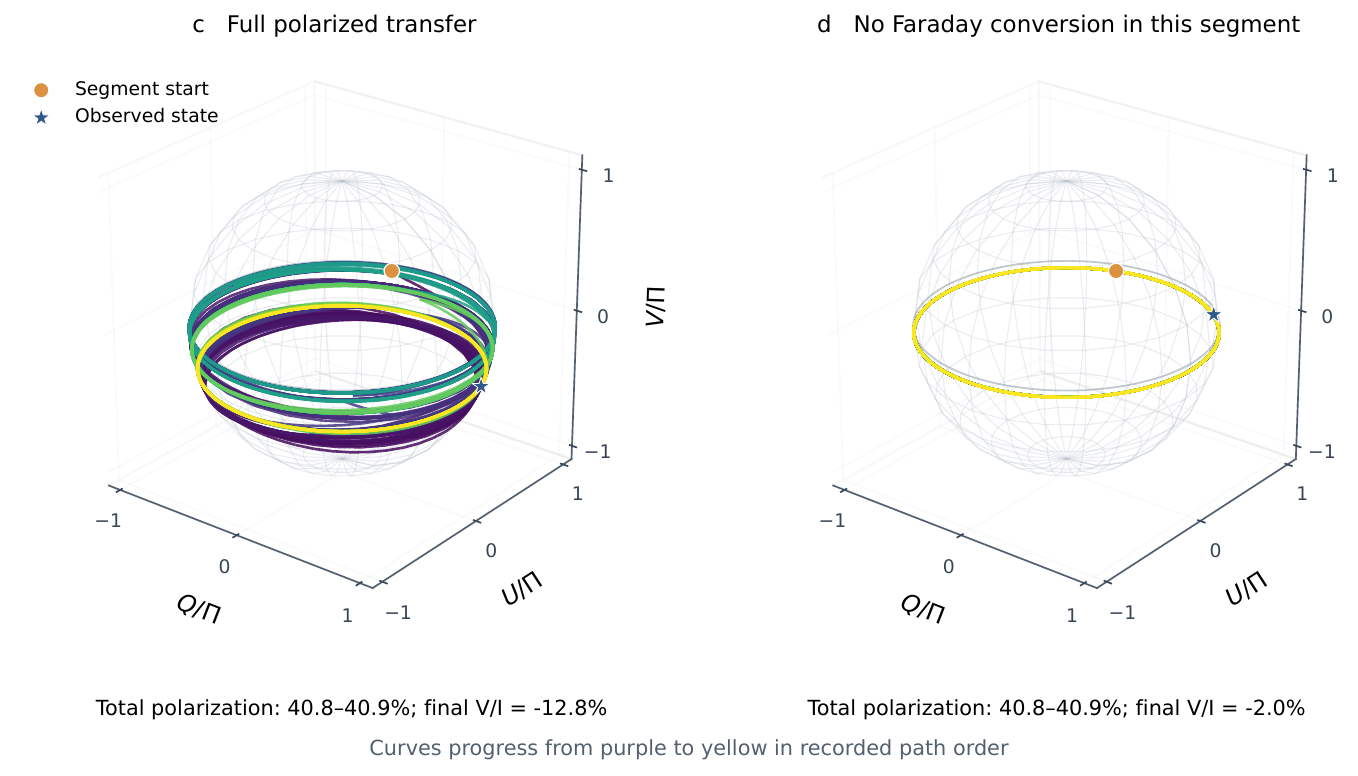}{Polarization evolution on the selected segment.}
 \end{minipage}
 \caption{Selected-ray analysis at 86 GHz and $i=85^\circ$.
(a) Stokes $I$, EVPA ticks, and the selected pixel.
(b) Its geodesic in Cartesian Kerr--Schild coordinates, with a density
slice for context. Magenta and teal intervals locate intrinsic circular
emission and Faraday conversion using the cumulative-contribution
criterion in the text. The arrow points toward the observer; the orange
point starts the segment shown in (c, d).
(c) Full transfer and (d) transfer without conversion on that segment,
starting from identical Stokes vectors. The Poincar\'e spheres show
$(Q,U,V)/\Pi$ in a parallel-transported basis, with
$\Pi=(Q^2+U^2+V^2)^{1/2}$. Orange circles and blue stars mark the start
and final states; colors progress from purple to yellow along the ray.
The total polarization fraction $\Pi/I$ and final $V/I$ appear below
each sphere.}
 \label{fig:diagnostic-example}
\end{figure*}

Figure~\ref{fig:diagnostic-example} follows a selected ray through a SANE snapshot at 86 GHz and $i=85^\circ$. Intrinsic circular emission is concentrated in a compact region, followed by more extended Faraday conversion. The highlighted intervals span 5\%--95\% of each process's cumulative absolute contribution to final Stokes $V$, locating where it acts most strongly. Opposite-sign contributions can cancel, so these intervals do not enclose 90\% of the net circular polarization. Their construction is specified in Appendix~\ref{app:diagnostic-checks}.

Panels (c) and (d) examine an outgoing segment where the invariant intensity has approached its final value. Suppressing conversion on this segment ($\rho_Q=\rho_U=0$), with the same incident Stokes vector and remaining coefficients, changes the final $V/I$ from $-12.8\%$ to $-2.0\%$. The total polarization fraction remains within $40.8\%$--$40.9\%$ in both calculations. Thus propagation after most of the intensity has accumulated can substantially change the balance between linear and circular polarization \citep{JonesOdell1977,Ricarte2021}. Ray histories locate this evolution and allow its effect on the emerging radiation to be tested.

\section{Conclusions}\label{sec:conclusions}

\kpolaris{} combines CPU/GPU polarized imaging with diagnostics of emission and propagation. Two-pass ray integration and snapshot streaming support multi-frequency and time-dependent calculations with bounded trajectory and input storage.

Analytic tests and refinement studies verify the numerical accuracy,
while comparisons with ipole show close agreement in all four Stokes
components for both fast-light and slow-light calculations. At separately assessed performance settings, complete $256^2$ fast-light images take a few seconds on the workstation and server GPUs. Slow-light batches on the RTX 5090 reach $2.47$ s per image with 15.1 GiB of peak memory, and a minimum-residency configuration operates at 5.4 GiB memory. These results demonstrate that slow-light image sequences can be generated efficiently on a single GPU with a configurable memory footprint.

Plasma maps, emission contributions, ray histories, and parameter sensitivities connect these images to the underlying flow and transfer processes. The selected-ray example shows how Faraday conversion can substantially alter circular polarization after most of the intensity has accumulated. These diagnostics help identify where radiation originates, how propagation modifies its polarization, and how the observed Stokes parameters respond to changes in the plasma.
Together with GPU-accelerated imaging, they provide a practical tool for interpreting polarized black-hole images and exploring their dependence on physical models across observing frequencies and time.

\section*{Data and Software}
\kpolaris{} is publicly available under the BSD-3-Clause
license at \url{https://github.com/zelinzh/KPolaris}.
The repository includes documentation and examples for running polarized radiative-transfer calculations. An archived software
release is available through Zenodo \citep{zhang_2026_22879728}.

\begin{acknowledgments}
We thank Zhenyu Zhang for helpful discussions, and thank Akhil Uniyal and Indu K. Dihingia for providing partial data support for code testing. The work is partly supported by NSFC Grant No. 12275004, 12547132 and No. 12588101.
\end{acknowledgments}

\software{KPolaris \citep{zhang_2026_22879728}, Dense Dump Codec \citep{zhang_2026_22845273}, Kokkos
\citep{Edwards2014,Trott2022}, HDF5 \citep{HDF5}, NumPy, h5py, Matplotlib,
ipole \citep{MoscibrodzkaGammie2018}}

\clearpage
\appendix

\section{Numerical Controls}\label{app:controls}

Table~\ref{tab:integration-controls} lists the integration settings for the
code comparisons and their refinement checks. 
For the slow-light refinement test, the largest component NMSE
between the coarser and refined \kpolaris{} images is $4.70\times10^{-4}$,
with the refined image used as the reference.

\begin{deluxetable*}{llrrrrr}[!t]
\tablecaption{Integration settings}\label{tab:integration-controls}
\tablewidth{0pt}
\tabletypesize{\footnotesize}
\tablehead{\colhead{Calculation} & \colhead{Model} &
\colhead{Tolerance} & \colhead{Geometry step} & \colhead{Radiation step} &
\colhead{Absorption depth} & \colhead{Faraday depth}}
\startdata
Fast-light comparison & RIAF & $10^{-12}$ & 0.5 & 0.25 & 1 & 4 \\
 & KHARMA & $10^{-15}$ & 0.25 & 0.125 & 0.5 & 2 \\
 & AthenaK & $10^{-15}$ & 0.03125 & 0.001953125 & 0.0625 & 0.25 \\
Slow light, refined & KHARMA & $10^{-13}$ & 5 & 0.0625 & 1 & 4 \\
Slow light, coarser & KHARMA & $10^{-12}$ & 5 & 0.125 & 1 & 4 \\
\enddata
\tablecomments{Tolerance is the requested local geometrical tolerance.
The step and depth columns give upper limits. Both step limits refer to increments in the affine parameter $\lambda$, using the normalization adopted in the code. Optical and Faraday depths are dimensionless. The slow-light refined/coarser ipole calculations use $\epsilon=5\times10^{-4}/10^{-3}$, respectively. 
Both slow-light refinement levels use a GPU snapshot-cache budget
of 13.5 GiB and retain the native input cadence.
Fast-light CPU/GPU performance comparisons use the settings in
Table~\ref{tab:daily-performance}. The slow-light throughput settings
are given in Appendix~\ref{app:performance-setup}.}
\end{deluxetable*}

Table~\ref{tab:memory} records memory use for the fast-light performance
tests and the slow-light comparison with ipole at the refined accuracy
settings in Table~\ref{tab:integration-controls}. The slow-light
\kpolaris{} run has a measured peak GPU memory use of 14.61 GiB for the
selected 13.5 GiB snapshot-cache budget. This peak reflects the selected validation configuration, not a minimum
algorithmic requirement. Memory use for the separate slow-light throughput tests
is reported with their timings in Table~\ref{tab:slowlight-throughput}.

\begin{deluxetable*}{llrr}[!t]
\tablecaption{Measured memory use}\label{tab:memory}
\tablewidth{0pt}
\tabletypesize{\footnotesize}
\tablehead{\colhead{Calculation} & \colhead{Resolution} &
\colhead{Host (GiB)} & \colhead{GPU (GiB)}}
\startdata
Fast light, GPU execution & $256^2$ & 1.62 & 3.18 \\
Fast light, largest GPU images & $2048^2$ & 1.62 & 4.25 \\
Fast light, GPU 1 & up to $1024^2$ & 1.61 & 2.88 \\
Slow light, \kpolaris{} & $256^2$ & 1.01 & 14.61 \\
Slow light, ipole & $256^2$ & 226.4 & \nodata \\
\enddata
\tablecomments{Values are maxima over the indicated calculations.
Host memory is the process resident-set high-water mark, except for the
ipole slow-light runs, where the maximum sampled resident plus swapped
memory is reported. Host RSS excludes the system file cache. For the slow-light
\kpolaris{} run, GPU memory is the sampled total device memory use during
exclusive execution, including the CUDA context. Fast-light GPU values
remain sampled process memory, including the CUDA context. Fast-light runs use
Table~\ref{tab:daily-performance}, with no observed swap use. Slow-light
values refer to the refined accuracy comparison in
Table~\ref{tab:integration-controls}.}
\end{deluxetable*}

\section{Simulation Samples and Cameras}\label{app:samples}

Table~\ref{tab:comparison-inputs} specifies the models and cameras used for
the ipole comparisons. The public KHARMA snapshots are from the v3 library
of \citet{Dhruv2025}. The slow-light sequence is a separate MAD simulation, and its native snapshot timestamps are used for temporal interpolation.

\begin{deluxetable*}{llrrrr}[!t]
\tablecaption{Inputs for the image comparisons}\label{tab:comparison-inputs}
\tablewidth{0pt}
\tabletypesize{\footnotesize}
\tablehead{\colhead{Model} & \colhead{Approximation} & \colhead{Spin $a$} &
\colhead{Simulation time ($M$)} & \colhead{$i$ (deg)} & \colhead{Field of view ($\mu$as)}}
\startdata
RIAF & Fast light & 0.9375 & \nodata & 85 & 200 \\
KHARMA MAD & Fast light & 0.9375 & 27500 & 17 & 160 \\
AthenaK MAD & Fast light & 0.98 & 25000 & 17 & 160 \\
KHARMA MAD & Slow light & 0.5 & 29000--29300 & 17 & 200 \\
\enddata
\tablecomments{All comparisons use $256^2$ pixels, 230 and 345 GHz. KHARMA uses a
$288\times128\times128$ FMKS grid with $\Gamma=4/3$ and the iharm/pyharm HDF5
schema. AthenaK uses Cartesian Kerr--Schild coordinates, seven static mesh
refinement levels, and $\Gamma=5/3$. The slow-light sequence spans $300M$
with cadence approximately $0.1M$, covering the roughly $260M$ range of
retarded times sampled by the rays.}
\end{deluxetable*}

\section{Performance Test Configurations}
\label{app:performance-setup}

Table~\ref{tab:hardware} specifies the processors used for the fast-light comparisons. Table~\ref{tab:daily-performance} records the corresponding integration controls and representative image errors on GPU 2. These settings are distinct from the tighter controls used for the independent-code accuracy comparisons in Table~\ref{tab:integration-controls}.

\begin{deluxetable*}{llcl}[!t]
\tablecaption{Processor groups used in the performance tests}\label{tab:hardware}
\tablewidth{0pt}
\tabletypesize{\footnotesize}
\tablehead{\colhead{Group} & \colhead{CPU} & \colhead{Cores/threads} & \colhead{GPU}}
\startdata
1: Laptop & AMD Ryzen 7 5800H & 8/16 & NVIDIA GeForce RTX 3060 Laptop (6 GiB) \\
2: Workstation & AMD Ryzen 9 9950X3D & 16/32 & NVIDIA GeForce RTX 5090 (32 GiB) \\
3: Server & AMD EPYC 9754 & 128/256 & NVIDIA A100-SXM4-40GB (40 GiB) \\
\enddata
\tablecomments{CPU $n$ and GPU $n$ denote the processors in group $n$ throughout the fast-light platform comparisons. The multithreaded calculations use all available hardware threads.}
\end{deluxetable*}

\begin{deluxetable*}{lrrrr}[!t]
\tablecaption{Fast-light accuracy and timing at the adopted performance settings}\label{tab:daily-performance}
\tablewidth{0pt}
\tabletypesize{\footnotesize}
\tablehead{\colhead{Model} & \colhead{GPU integration (s)} &
\colhead{Complete image (s)} & \colhead{NMSE(I)} &
\colhead{$\max_{I,Q,U,V}\mathrm{NMSE}$}}
\startdata
RIAF & 1.85 & 2.13 & $1.43\times10^{-8}$ & $1.53\times10^{-7}$ \\
KHARMA & 2.75 & 3.19 & $3.10\times10^{-4}$ & $1.66\times10^{-3}$ \\
AthenaK & 3.06 & 5.09 & $6.71\times10^{-4}$ & $1.79\times10^{-3}$ \\
\enddata
\tablecomments{Geometrical tolerance $10^{-10}$, maximum geometrical and radiation steps $10$ and $0.25$, and absorption/Faraday-depth limits $1$ and $4$. Values are medians of three runs after a warm-up.
Integration time includes geodesics and polarized transfer. Complete image time
also includes initialization, input preparation, data transfer, and HDF5 output,
with warm file caches. NMSE is measured against a more finely integrated
\kpolaris{} image of the same model.}
\end{deluxetable*}

The slow-light benchmark in Section~\ref{sec:slowlight-performance}
uses the native $288\times128\times128$ KHARMA grid, with approximately
82 GB of local input per 16-image batch. The snapshot spacing is nominally
$0.5M$ and the image spacing is $0.5M$, distinct from the native
approximately $0.1M$ cadence of the slow-light ipole comparison.
Calculations use double precision, with geometrical
tolerance $10^{-10}$, maximum geometrical and radiation steps $5$ and
$0.25$, and absorption/Faraday-depth limits $1$ and $4$.
The minimum-residency configuration uses one snapshot window without
upload double buffering. The intermediate configuration enables pipelined
uploads. Peak GPU memory is the sampled total device memory use, with
no other GPU processes running. 

% The first-frame accuracy reference uses the same snapshots, geometrical tolerance $10^{-12}$, and maximum radiation step $0.0625$, retaining maximum geometrical step $5$. It is excluded from the timing statistics.

\section{Auxiliary Quantities and Emission Contributions}
\label{app:diagnostic-definitions}

For a local plasma quantity $X$, the emissivity-weighted mean is
\begin{equation}
 \langle X\rangle_{\mathcal J_I}
 =\frac{\int X\mathcal J_I\,d\lambda}{\int\mathcal J_I\,d\lambda}.
\end{equation}
The code's affine normalization gives $\mathcal J_I=j_{\nu,I}/\nu^2$ and $d\ell=\nu\,d\lambda$. Thus $\mathcal J_I\,d\lambda=j_{\nu,I}\,d\ell/\nu^3$ includes frequency shifts but excludes absorption and Faraday propagation. The absorption optical depth is $\tau_I=\int\nu\alpha_{\nu,I}\,d\lambda
=\int\alpha_{\nu,I}\,d\ell$.
The code stores means of radius, density, electron temperature, magnetic field strength, and magnetization, together with optical and Faraday depths. These outputs are analogous to auxiliary maps described by \citet{White2022}.  The accumulated Faraday rotation coefficient $\int\mathcal R_V\,d\lambda$ and the accumulated magnitude of the Faraday coefficients, $\int(\mathcal R_Q^2+\mathcal R_U^2+\mathcal R_V^2)^{1/2}\,d\lambda$, are propagation integrals.  Neither is generally the observed EVPA change in an emitting, absorbing plasma with conversion.

Emission contributions instead retain the intervening transfer. Partitioning the source into regions $k$, we integrate
\begin{equation}\label{eq:radial-contribution}
 \frac{d\boldsymbol{\mathcal S}^{(k)}}{d\lambda}
 =\mathbf1_k\boldsymbol{\mathcal J}-\mathbf K\boldsymbol{\mathcal S}^{(k)},
 \qquad \boldsymbol{\mathcal S}^{(k)}(\lambda_0)=0.
\end{equation}
For a complete partition and zero incident radiation, linearity gives
$\sum_k\boldsymbol{\mathcal S}^{(k)}_{\rm obs}=
\boldsymbol{\mathcal S}_{\rm obs}$.
The code supports radial shells, geometric regions, and ranges of plasma
variables.  The contributions identify the seed emission and its conversion into $V$ can occur elsewhere along the ray.

At each pixel $p$, the relative $L_1$ source-sum residual is
\begin{equation}
 \epsilon_p =
 \frac{\displaystyle
   \sum_{X\in\{I,Q,U,V\}}
   \left|\sum_k \mathcal S^{(k)}_{X,p}
                   -\mathcal S_{X,p}\right|}
 {\displaystyle
   \sum_{X\in\{I,Q,U,V\}}|\mathcal S_{X,p}|},
\end{equation}
where all Stokes quantities are evaluated at the observer and the denominator is nonzero. Independent checks of 27 archived $256^2$ contribution arrays give $\max_p\epsilon_p<1.62\times10^{-13}$, verifying closure of the regional emission decomposition.

\section{Ray Transfer and Parameter Sensitivities}
\label{app:diagnostic-checks}

\subsection{Selected-ray Display and Transfer Checks}\label{app:cp-accounting}

The displayed outgoing segment begins at the first recorded point after
which the invariant intensity $\mathcal I$ remains within 1\% of its final
value. Both replays start from the same Stokes vector, and the modified
replay suppresses conversion only on this segment. This finite change
in the transfer differs from a parameter derivative about the original
model.

The highlighted intervals use the $\mathcal J_V$ and
$\mathcal R_Q\mathcal U-\mathcal R_U\mathcal Q$ terms evaluated along
the full solution. Their absolute stepwise contributions, weighted by
subsequent scalar absorption, are accumulated in forward path order.
The first samples reaching 5\% and 95\% define the endpoints. The display
segment and contribution intervals are selected independently.

An independent replay with absorption-matrix diagonalization
and Rodrigues rotations reproduces the terminal Stokes vector to
$1.65\times10^{-13}$ in relative L1 across three test rays.
Two and four subdivisions per interval change the selected ray's circular
fraction by at most $1.2\times10^{-4}$ and preserve the coefficient-removal
result. The displayed Poincar\'e curves use 16 subdivisions.

\subsection{Derivative Checks}
\label{app:response-checks}

We check the sensitivities defined in Section~\ref{sec:response-definitions} against complete transfer solutions at logarithmic parameter offsets $q=\pm h$. For each Stokes component, the centered difference is
\begin{equation}\label{eq:response-finite-difference}
 \boldsymbol{\mathcal D}_h=
 \frac{\boldsymbol{\mathcal S}(h)-\boldsymbol{\mathcal S}(-h)}{2h}.
\end{equation}
Repeating the calculation at $q=\pm h/2$ tests the parameter-step truncation error. In all perturbations, the ray geometry, velocity, and baseline emitting-domain mask are held fixed, as in the sensitivity calculation.

For the $400^2$ MAD sensitivity test, halving the geometrical and radiation step caps and the absorption and Faraday depth caps changes the full-Stokes derivatives by $0.223\%$, $0.114\%$, and $0.184\%$ in relative L1 for density
normalization, electron temperature, and magnetic field strength, respectively. At fixed ray sampling, the sums of the emission, absorption, rotation, and conversion responses agree with the complete transfer differences evaluated at half the parameter step to better than $4.2\times10^{-8}$. 

\bibliographystyle{aasjournalv7.1}
\bibliography{references}

@article{JonesOdell1977,
  author = {Jones, T. W. and O'Dell, S. L.},
  title = {Transfer of polarized radiation in self-absorbed synchrotron sources. {I}. Results for a homogeneous source},
  journal = {The Astrophysical Journal},
  year = {1977},
  volume = {214},
  pages = {522--539},
  doi = {10.1086/155278}
}

@article{Strang1968,
  author = {Strang, Gilbert},
  title = {On the Construction and Comparison of Difference Schemes},
  journal = {SIAM Journal on Numerical Analysis},
  year = {1968},
  volume = {5},
  number = {3},
  pages = {506--517},
  doi = {10.1137/0705041}
}

@article{MoscibrodzkaGammie2018,
  author = {{Mo\'scibrodzka}, Monika and Gammie, Charles F.},
  title = {ipole: semi-analytic scheme for relativistic polarized radiative transport},
  journal = {Monthly Notices of the Royal Astronomical Society},
  year = {2018},
  volume = {475},
  number = {1},
  pages = {43--54},
  doi = {10.1093/mnras/stx3162},
  eprint = {1712.03057},
  archivePrefix = {arXiv}
}

@article{MoscibrodzkaYfantis2023,
  author = {{Mo\'scibrodzka}, Monika A. and Yfantis, Aristomenis I.},
  title = {Prospects for Ray-tracing Light Intensity and Polarization in Models of Accreting Compact Objects Using a {GPU}},
  journal = {The Astrophysical Journal Supplement Series},
  year = {2023},
  volume = {265},
  number = {1},
  pages = {22},
  doi = {10.3847/1538-4365/acb6f9},
  eprint = {2302.02733},
  archivePrefix = {arXiv}
}

@article{Prather2023,
  author = {Prather, Ben S. and Dexter, Jason and Mo{\'s}cibrodzka, Monika and Pu, Hung-Yi and Bronzwaer, Thomas and Davelaar, Jordy and Younsi, Ziri and Gammie, Charles F. and Gold, Roman and Wong, George N. and others},
  title = {Comparison of Polarized Radiative Transfer Codes Used by the EHT Collaboration},
  journal = {The Astrophysical Journal},
  year = {2023},
  volume = {950},
  number = {1},
  eid = {35},
  doi = {10.3847/1538-4357/acc586},
  eprint = {2303.12004},
  archivePrefix = {arXiv}
}

@article{Pihajoki2018,
  author = {Pihajoki, Pauli and Mannerkoski, Matias and N\"attil\"a, Joonas and Johansson, Peter H.},
  title = {General Purpose Ray-Tracing and Polarized Radiative Transfer in General Relativity},
  journal = {The Astrophysical Journal},
  year = {2018},
  volume = {863},
  number = {1},
  eid = {8},
  doi = {10.3847/1538-4357/aacea0},
  eprint = {1804.04670},
  archivePrefix = {arXiv}
}

@article{White2022,
  author = {White, Christopher J.},
  title = {Blacklight: A General-relativistic Ray-tracing and Analysis Tool},
  journal = {The Astrophysical Journal Supplement Series},
  year = {2022},
  volume = {262},
  number = {1},
  pages = {28},
  doi = {10.3847/1538-4365/ac77ef},
  eprint = {2203.15963},
  archivePrefix = {arXiv}
}

@article{Dexter2016,
  author = {Dexter, Jason},
  title = {A public code for general relativistic, polarised radiative transfer around spinning black holes},
  journal = {Monthly Notices of the Royal Astronomical Society},
  year = {2016},
  volume = {462},
  number = {1},
  pages = {115--136},
  doi = {10.1093/mnras/stw1526},
  eprint = {1602.03184},
  archivePrefix = {arXiv}
}

@article{Pandya2016,
  author = {Pandya, Alex and Zhang, Zhaowei and Chandra, Mani and Gammie, Charles F.},
  title = {Polarized Synchrotron Emissivities and Absorptivities for Relativistic Thermal, Power-law, and Kappa Distribution Functions},
  journal = {The Astrophysical Journal},
  year = {2016},
  volume = {822},
  number = {1},
  eid = {34},
  doi = {10.3847/0004-637X/822/1/34},
  eprint = {1602.08749},
  archivePrefix = {arXiv}
}

@article{Edwards2014,
  author = {Edwards, H. Carter and Trott, Christian R. and Sunderland, Daniel},
  title = {Kokkos: Enabling manycore performance portability through polymorphic memory access patterns},
  journal = {Journal of Parallel and Distributed Computing},
  year = {2014},
  volume = {74},
  number = {12},
  pages = {3202--3216},
  doi = {10.1016/j.jpdc.2014.07.003}
}

@article{Trott2022,
  author = {Trott, Christian R. and Lebrun-Grandi\'e, Damien and Arndt, Daniel and Ciesko, Jan and Dang, Vinh and Ellingwood, Nathan and Gayatri, Rahulkumar and Harvey, Evan and Hollman, Daisy S. and Ibanez, Dan and Liber, Nevin and Madsen, Jonathan and Miles, Jeff and Poliakoff, David and Powell, Amy and Rajamanickam, Sivasankaran and Simberg, Mikael and Sunderland, Dan and Turcksin, Bruno and Wilke, Jeremiah},
  title = {Kokkos 3: Programming Model Extensions for the Exascale Era},
  journal = {IEEE Transactions on Parallel and Distributed Systems},
  year = {2022},
  volume = {33},
  number = {4},
  pages = {805--817},
  doi = {10.1109/TPDS.2021.3097283}
}

@article{Dhruv2025,
  author = {Dhruv, Vedant and Prather, Ben and Wong, George N. and Gammie, Charles F.},
  title = {A Survey of General Relativistic Magnetohydrodynamic Models for Black Hole Accretion Systems},
  journal = {The Astrophysical Journal Supplement Series},
  year = {2025},
  volume = {277},
  number = {1},
  eid = {16},
  doi = {10.3847/1538-4365/adaea6},
  eprint = {2411.12647},
  archivePrefix = {arXiv}
}

@misc{HDF5,
  author = {{The HDF Group}},
  title = {Hierarchical Data Format, version 5},
  url = {https://www.hdfgroup.org/solutions/hdf5/},
  year = {2026}
}

@article{Ricarte2020,
  author = {Ricarte, Angelo and Prather, Ben S. and Wong, George N. and Narayan, Ramesh and Gammie, Charles and Johnson, Michael D.},
  title = {Decomposing the internal Faraday rotation of black hole accretion flows},
  journal = {Monthly Notices of the Royal Astronomical Society},
  year = {2020}, volume = {498}, number = {4}, pages = {5468--5488},
  doi = {10.1093/mnras/staa2692}, eprint = {2009.02369}, archivePrefix = {arXiv}
}

@article{Ricarte2021,
  author = {Ricarte, Angelo and Qiu, Richard and Narayan, Ramesh},
  title = {Black hole magnetic fields and their imprint on circular polarization images},
  journal = {Monthly Notices of the Royal Astronomical Society},
  year = {2021}, volume = {505}, number = {1}, pages = {523--539},
  doi = {10.1093/mnras/stab1289}
}

@article{Motta2025,
  author = {Motta, Pedro Naethe and Prather, Ben S. and C{\'a}rdenas-Avenda{\~n}o, Alejandro},
  title = {{Jipole}: A Differentiable {ipole}-based Code for Radiative Transfer in Curved Spacetimes},
  journal = {The Astrophysical Journal},
  year = {2025}, volume = {995}, eid = {56},
  doi = {10.3847/1538-4357/ae16a0},
  eprint = {2509.07065}, archivePrefix = {arXiv}
}

@article{Motta2026,
  author = {Motta, Pedro Naethe and Raia Neto, M{\'a}rio and Prather, Cora and C{\'a}rdenas-Avenda{\~n}o, Alejandro},
  title = {Sensitivities of Black Hole Images from {GRMHD} Simulations},
  journal = {The Astrophysical Journal},
  year = {2026}, volume = {1004}, eid = {218},
  doi = {10.3847/1538-4357/ae733f},
  eprint = {2604.11869}, archivePrefix = {arXiv}
}

@article{Wong2022,
  author = {Wong, George N. and Prather, Ben S. and Dhruv, Vedant and Ryan, Benjamin R. and Mo{\'s}cibrodzka, Monika and Chan, Chi-kwan and Joshi, Abhishek V. and Yarza, Ricardo and Ricarte, Angelo and Shiokawa, Hotaka and Dolence, Joshua C. and Noble, Scott C. and McKinney, Jonathan C. and Gammie, Charles F.},
  title = {{PATOKA}: Simulating Electromagnetic Observables of Black Hole Accretion},
  journal = {The Astrophysical Journal Supplement Series},
  year = {2022},
  volume = {259},
  eid = {64},
  doi = {10.3847/1538-4365/ac582e},
  eprint = {2202.11721},
  archivePrefix = {arXiv}
}

@article{Younsi2020,
  author = {Younsi, Ziri and Porth, Oliver and Mizuno, Yosuke and Fromm, Christian M. and Olivares, Hector},
  title = {Modelling the polarised emission from black holes on event horizon-scales},
  journal = {Proceedings of the International Astronomical Union},
  year = {2020},
  volume = {14},
  number = {S342},
  pages = {9--12},
  doi = {10.1017/S1743921318007263},
  eprint = {1907.09196},
  archivePrefix = {arXiv}
}

@article{Bronzwaer2020,
  author = {Bronzwaer, Thomas and Younsi, Ziri and Davelaar, Jordy and Falcke, Heino},
  title = {{RAPTOR}. {II}. Polarized radiative transfer in curved spacetime},
  journal = {Astronomy \& Astrophysics},
  year = {2020},
  volume = {641},
  pages = {A126},
  doi = {10.1051/0004-6361/202038573},
  eprint = {2007.03045},
  archivePrefix = {arXiv}
}

@article{Aimar2024,
  author = {Aimar, N. and Paumard, T. and Vincent, F. H. and Gourgoulhon, E. and Perrin, G.},
  title = {{GYOTO} 2.0: a polarized relativistic ray-tracing code},
  journal = {Classical and Quantum Gravity},
  year = {2024},
  volume = {41},
  number = {9},
  pages = {095010},
  doi = {10.1088/1361-6382/ad351d},
  eprint = {2311.18802},
  archivePrefix = {arXiv}
}

@article{Huang2024,
  author = {Huang, Jiewei and Zheng, Liheng and Guo, Minyong and Chen, Bin},
  title = {{Coport}: a new public code for polarized radiative transfer in a covariant framework},
  journal = {Journal of Cosmology and Astroparticle Physics},
  year = {2024},
  volume = {2024},
  number = {11},
  pages = {054},
  doi = {10.1088/1475-7516/2024/11/054},
  eprint = {2407.10431},
  archivePrefix = {arXiv}
}

@article{Chan2013,
  author = {Chan, Chi-kwan and Psaltis, Dimitrios and {\"O}zel, Feryal},
  title = {{GRay}: A Massively Parallel {GPU}-Based Code for Ray Tracing in Relativistic Spacetimes},
  journal = {The Astrophysical Journal},
  year = {2013},
  volume = {777},
  number = {1},
  eid = {13},
  doi = {10.1088/0004-637X/777/1/13},
  eprint = {1303.5057},
  archivePrefix = {arXiv}
}

@article{Pu2016,
  author = {Pu, Hung-Yi and Yun, Kiyun and Younsi, Ziri and Yoon, Suk-Jin},
  title = {{Odyssey}: A Public {GPU}-Based Code for General Relativistic Radiative Transfer in {Kerr} Spacetime},
  journal = {The Astrophysical Journal},
  year = {2016},
  volume = {820},
  number = {2},
  eid = {105},
  doi = {10.3847/0004-637X/820/2/105},
  eprint = {1601.02063},
  archivePrefix = {arXiv}
}

@article{Bronzwaer2018,
  author = {Bronzwaer, Thomas and Davelaar, Jordy and Younsi, Ziri and Mo{\'s}cibrodzka, Monika and Falcke, Heino and Kramer, Michael and Rezzolla, Luciano},
  title = {{RAPTOR}. {I}. Time-dependent radiative transfer in arbitrary spacetimes},
  journal = {Astronomy \& Astrophysics},
  year = {2018},
  volume = {613},
  eid = {A2},
  doi = {10.1051/0004-6361/201732149},
  eprint = {1801.10452},
  archivePrefix = {arXiv}
}

@article{EHTC2019,
  author = {{Event Horizon Telescope Collaboration} and others},
  title = {First {M87} Event Horizon Telescope Results. {I}. The Shadow of the Supermassive Black Hole},
  journal = {The Astrophysical Journal Letters}, year = {2019}, volume = {875}, eid = {L1},
  doi = {10.3847/2041-8213/ab0ec7}, eprint = {1906.11238}, archivePrefix = {arXiv}
}

@article{EHTC2021VII,
  author = {{Event Horizon Telescope Collaboration} and others},
  title = {First {M87} Event Horizon Telescope Results. {VII}. Polarization of the Ring},
  journal = {The Astrophysical Journal Letters}, year = {2021}, volume = {910}, eid = {L12},
  doi = {10.3847/2041-8213/abe71d}, eprint = {2105.01169}, archivePrefix = {arXiv}
}

@article{EHTC2021VIII,
  author = {{Event Horizon Telescope Collaboration} and others},
  title = {First {M87} Event Horizon Telescope Results. {VIII}. Magnetic Field Structure near The Event Horizon},
  journal = {The Astrophysical Journal Letters}, year = {2021}, volume = {910}, eid = {L13},
  doi = {10.3847/2041-8213/abe4de}, eprint = {2105.01173}, archivePrefix = {arXiv}
}

@article{EHTC2022,
  author = {{Event Horizon Telescope Collaboration} and others},
  title = {First Sagittarius {A*} Event Horizon Telescope Results. {I}. The Shadow of the Supermassive Black Hole in the Center of the Milky Way},
  journal = {The Astrophysical Journal Letters}, year = {2022}, volume = {930}, eid = {L12},
  doi = {10.3847/2041-8213/ac6674}, eprint = {2311.08680}, archivePrefix = {arXiv}
}

@article{EHTC2024VII,
  author = {{Event Horizon Telescope Collaboration} and others},
  title = {First Sagittarius {A*} Event Horizon Telescope Results. {VII}. Polarization of the Ring},
  journal = {The Astrophysical Journal Letters}, year = {2024}, volume = {964}, eid = {L25},
  doi = {10.3847/2041-8213/ad2df0}
}

@article{EHTC2024VIII,
  author = {{Event Horizon Telescope Collaboration} and others},
  title = {First Sagittarius {A*} Event Horizon Telescope Results. {VIII}. Physical Interpretation of the Polarized Ring},
  journal = {The Astrophysical Journal Letters}, year = {2024}, volume = {964}, eid = {L26},
  doi = {10.3847/2041-8213/ad2df1}
}

@article{Johnson2023,
  author = {Johnson, Michael D. and Akiyama, Kazunori and Blackburn, Lindy and Bouman, Katherine L. and Broderick, Avery E. and Cardoso, Vitor and Fender, R. P. and Fromm, Christian M. and Galison, Peter and G{\'o}mez, Jos{\'e} L. and Haggard, Daryl and Lister, Matthew L. and Lobanov, Andrei P. and Markoff, Sera and Narayan, Ramesh and Natarajan, Priyamvada and Nichols, Tiffany and Pesce, Dominic W. and Younsi, Ziri and Chael, Andrew and Chatterjee, Koushik and Chaves, Ryan and Doboszewski, Juliusz and Dodson, Richard and Doeleman, Sheperd S. and Elder, Jamee and Fitzpatrick, Garret and Haworth, Kari and Houston, Janice and Issaoun, Sara and Kovalev, Yuri Y. and Levis, Aviad and Lico, Rocco and Marcoci, Alexandru and Martens, Niels C. M. and Nagar, Neil M. and Oppenheimer, Aaron and Palumbo, Daniel C. M. and Ricarte, Angelo and Rioja, Mar{\'i}a J. and Roelofs, Freek and Thresher, Ann C. and Tiede, Paul and Weintroub, Jonathan and Wielgus, Maciek},
  title = {Key Science Goals for the Next-Generation Event Horizon Telescope},
  journal = {Galaxies}, year = {2023}, volume = {11}, number = {3}, eid = {61},
  doi = {10.3390/galaxies11030061}, eprint = {2304.11188}, archivePrefix = {arXiv}
}

@article{Gammie2003,
  author = {Gammie, Charles F. and McKinney, Jonathan C. and T{\'o}th, G{\'a}bor},
  title = {{HARM}: A Numerical Scheme for General Relativistic Magnetohydrodynamics},
  journal = {The Astrophysical Journal},
  year = {2003},
  volume = {589},
  number = {1},
  pages = {444--457},
  doi = {10.1086/374594},
  eprint = {astro-ph/0301509},
  archivePrefix = {arXiv}
}

@article{Komissarov2006,
  author = {Komissarov, S. S.},
  title = {Magnetized tori around {Kerr} black holes: analytic solutions with a toroidal magnetic field},
  journal = {Monthly Notices of the Royal Astronomical Society},
  year = {2006},
  volume = {368},
  pages = {993--1000},
  doi = {10.1111/j.1365-2966.2006.10183.x},
  eprint = {astro-ph/0601678},
  archivePrefix = {arXiv}
}

@article{Shcherbakov2008,
  author = {Shcherbakov, Roman V.},
  title = {Propagation Effects in Magnetized Transrelativistic Plasmas},
  journal = {The Astrophysical Journal}, year = {2008}, volume = {688}, number = {2},
  pages = {695--700}, doi = {10.1086/592326}, eprint = {0809.0012}, archivePrefix = {arXiv}
}

@article{Marszewski2021,
  author = {Marszewski, Andrew and Prather, Ben S. and Joshi, Abhishek V. and Pandya, Alex and Gammie, Charles F.},
  title = {Updated Transfer Coefficients for Magnetized Plasmas},
  journal = {The Astrophysical Journal}, year = {2021}, volume = {921}, number = {1},
  eid = {17}, doi = {10.3847/1538-4357/ac1b28}, eprint = {2108.10359}, archivePrefix = {arXiv}
}

@incollection{Prather2025,
  author = {Prather, Cora},
  title = {{KHARMA}: Flexible, Portable Performance for {GRMHD}},
  booktitle = {New Frontiers in {GRMHD} Simulations},
  series = {Springer Series in Astrophysics and Cosmology},
  publisher = {Springer}, year = {2025}, pages = {167--201},
  doi = {10.1007/978-981-97-8522-3_5}, eprint = {2408.01361}, archivePrefix = {arXiv}
}

@article{Prather2021,
  author = {Prather, Cora and Wong, George N. and Dhruv, Vedant and Ryan, Benjamin R. and Dolence, Joshua C. and Ressler, Sean M. and Gammie, Charles F.},
  title = {{iharm3D}: Vectorized General Relativistic Magnetohydrodynamics},
  journal = {Journal of Open Source Software}, year = {2021}, volume = {6}, number = {66},
  pages = {3336}, doi = {10.21105/joss.03336}
}

@article{Stone2026,
  author = {Stone, James M. and Mullen, Patrick D. and Fielding, Drummond and Grete, Philipp and Guo, Minghao and Kempski, Philipp and Most, Elias R. and White, Christopher J. and Wong, George N.},
  title = {{AthenaK}: A Performance-portable Version of the {Athena++} {AMR} Framework},
  journal = {The Astrophysical Journal Supplement Series}, year = {2026}, volume = {283},
  eid = {27}, doi = {10.3847/1538-4365/ae3717}, eprint = {2409.16053}, archivePrefix = {arXiv}
}

@article{Moscibrodzka2016,
  author = {Mo{\'s}cibrodzka, Monika and Falcke, Heino and Shiokawa, Hotaka},
  title = {General relativistic magnetohydrodynamical simulations of the jet in {M\,87}},
  journal = {Astronomy \& Astrophysics}, year = {2016}, volume = {586}, eid = {A38},
  doi = {10.1051/0004-6361/201526630}, eprint = {1510.07243}, archivePrefix = {arXiv}
}

@article{Broderick2016,
  author = {Broderick, Avery E. and Fish, Vincent L. and Johnson, Michael D. and others},
  title = {Modeling Seven Years of Event Horizon Telescope Observations with Radiatively Inefficient Accretion Flow Models},
  journal = {The Astrophysical Journal}, year = {2016}, volume = {820}, pages = {137},
  doi = {10.3847/0004-637X/820/2/137}
}

@article{PuBroderick2018,
  author = {Pu, Hung-Yi and Broderick, Avery E.},
  title = {Probing the Innermost Accretion Flow Geometry of Sagittarius {A*} with Event Horizon Telescope},
  journal = {The Astrophysical Journal}, year = {2018}, volume = {863}, pages = {148},
  doi = {10.3847/1538-4357/aad086}
}

@article{Gold2020,
  author = {Gold, Roman and Broderick, Avery E. and Younsi, Ziri and Fromm, Christian M. and Gammie, Charles F. and {Mo\'scibrodzka}, Monika and Pu, Hung-Yi and Bronzwaer, Thomas and Davelaar, Jordy and others},
  title = {Verification of Radiative Transfer Schemes for the {EHT}},
  journal = {The Astrophysical Journal},
  year = {2020},
  volume = {897},
  number = {2},
  eid = {148},
  doi = {10.3847/1538-4357/ab96c6}
}

@article{Porth2017,
  author = {Porth, Oliver and Olivares, Hector and Mizuno, Yosuke and Younsi, Ziri and Rezzolla, Luciano and {Mo\'scibrodzka}, Monika and Falcke, Heino and Kramer, Michael},
  title = {The black hole accretion code},
  journal = {Computational Astrophysics and Cosmology},
  year = {2017},
  volume = {4},
  number = {1},
  eid = {1},
  doi = {10.1186/s40668-017-0020-2},
  eprint = {1611.09720},
  archivePrefix = {arXiv}
}

@software{zhang_2026_22845273,
  author       = {Zhang, Zelin and
                  Zhang, Zhenyu and
                  Chen, Bin},
  title        = {Dense Dump Codec},
  month        = sep,
  year         = 2026,
  publisher    = {Zenodo},
  version      = {v1.0.0},
  doi          = {10.5281/zenodo.22845273},
  url          = {https://doi.org/10.5281/zenodo.22845273},
  swhid        = {swh:1:dir:c08e06ca04d3468f7f5bd4f3d486ae08ba57eb6c
                   ;origin=https://doi.org/10.5281/zenodo.22845272;vi
                   sit=swh:1:snp:a5593939a206ca496bd02d28bf605d7e0aca
                   acd4;anchor=swh:1:rel:d23ec28a8bc570a29b4a4b10cb93
                   58c323fe227b;path=zelinzh-dense-dump-codec-7a7e524
                  },
}

@software{zhang_2026_22879728,
  author       = {Zhang, Zelin and
                  Chen, Bin},
  title        = {KPolaris: GPU-accelerated Polarized Radiative
                   Transfer in General Relativity
                  },
  month        = sep,
  year         = 2026,
  publisher    = {Zenodo},
  version      = {v0.1.0},
  doi          = {10.5281/zenodo.22879728},
  url          = {https://doi.org/10.5281/zenodo.22879728},
}

\end{document}